\documentclass[aps,onecolumn,superscriptaddress,longbibliography,nofootinbib]{revtex4-2}

\usepackage{amsmath}
\usepackage{amssymb}
\usepackage{amstext}
\usepackage{amsopn}
\usepackage{amsfonts}
\usepackage{amsxtra}
\usepackage[english]{babel}
\usepackage{graphicx}
\usepackage{bm}
\usepackage{multirow}
\usepackage{dcolumn}
\usepackage{color}
\usepackage{xcolor}
\usepackage{verbatim}
\usepackage{lineno}
\usepackage{lipsum}
\usepackage{cuted}
\def\RuCl{$\alpha$-RuCl$_3$}

\begin{document}

\title{Divergence of Majorana-Phonon Scattering in Kitaev Quantum Spin Liquid}

\author{Haoxiang Li}
\affiliation{Materials Science and Technology Division, Oak Ridge National Laboratory, Oak Ridge, Tennessee 37831, USA}
\author{A. Said}
\affiliation{Advanced Photon Source, Argonne National Laboratory, Argonne, Illinois 60439, USA}
\author{J.Q. Yan}
\affiliation{Materials Science and Technology Division, Oak Ridge National Laboratory, Oak Ridge, Tennessee 37831, USA}
\author{D.M. Mandrus}
\affiliation{Materials Science and Technology Division, Oak Ridge National Laboratory, Oak Ridge, Tennessee 37831, USA}
\affiliation{Department of Materials Science and Engineering, University of Tennessee at Knoxville, Knoxville, Tennessee 37996, USA}
\author{H. N. Lee}
\affiliation{Materials Science and Technology Division, Oak Ridge National Laboratory, Oak Ridge, Tennessee 37831, USA}
\author{S. Okamoto}
\affiliation{Materials Science and Technology Division, Oak Ridge National Laboratory, Oak Ridge, Tennessee 37831, USA}
\author{G\'abor B. Hal\'asz}\email{halaszg@ornl.gov}
\affiliation{Materials Science and Technology Division, Oak Ridge National Laboratory, Oak Ridge, Tennessee 37831, USA}
\author{H. Miao}\email{miaoh@ornl.gov}
\affiliation{Materials Science and Technology Division, Oak Ridge National Laboratory, Oak Ridge, Tennessee 37831, USA}

\date{\today}

\begin{abstract}
Magnetoelastic interaction couples spin and lattice degrees of freedom and plays a key role in thermal transport properties of magnetic insulators. In the Kitaev quantum spin liquid, the low energy excitations are charge neutral Majorana fermions, which transform the magnetoelasctic interaction into Majorana-phonon scattering. Motivated by anomalous thermal properties of the Kitaev quantum spin liquid candidate \RuCl{}, in this letter, we combine meV resolution inelastic x-ray scattering and theoretical calculation to examine the Majorana-phonon scattering. We analytically derive the velocity-dependent Majorana-phonon scattering and find a divergence when the acoustic phonons and the itinerant Majorana fermions have the same velocity. Based on the experimentally determined acoustic phonon velocity in \RuCl{}, we estimate the range in the Kitaev interaction for which divergent Majorana-phonon scattering can happen. Our result opens a new avenue to uncover fractionalized quasiparticles in the Kitaev quantum spin liquid and emphasizes the critical role of lattice excitations in \RuCl{}.

\end{abstract}

\maketitle


Thermal transport is one of the key physical properties to uncover the nature of emergent quantum states, such as unconventional superconductivity~\cite{Ronning2006,Bourgeois2016, Grissonnanche2019, Samajdar2019, Han2019} and other exotic states that host anyonic excitations~\cite{Kitaev2006,Read2000, Nasu2016, Banerjee2017,Leahy2017, Banerjee2018, Hentrich2018, Kasahara2018, Yamashita2020, Yokoi2021,Czajka2021}. In the Kitaev honeycomb model \cite{Kitaev2006}, \begin{equation}
H=\sum_{\gamma,<i,j>}J^{\gamma}_K S^{\gamma}_i S^{\gamma}_j
\label{KitaevModel}
\end{equation}
(where $J^{\gamma}_K$ ($\gamma=X,Y,Z$) is the bond-dependent coupling parameter, and $<i,j>$ stands for nearest-neighbor pairs of spins at one of the $X$, $Y$, or $Z$ bonds), the low-energy spin excitations fractionalize into charge neutral Majorana fermions. Interestingly, theoretical studies found that Majorana fermions can couple with collective lattice excitations, known as phonons, through the magnetoelastic interaction and give rise to unusual thermal transport behaviors \cite{Ye2020, Feng2021, Nasu2016}. 

\begin{figure*}
\centering
\includegraphics[width=0.75\linewidth]{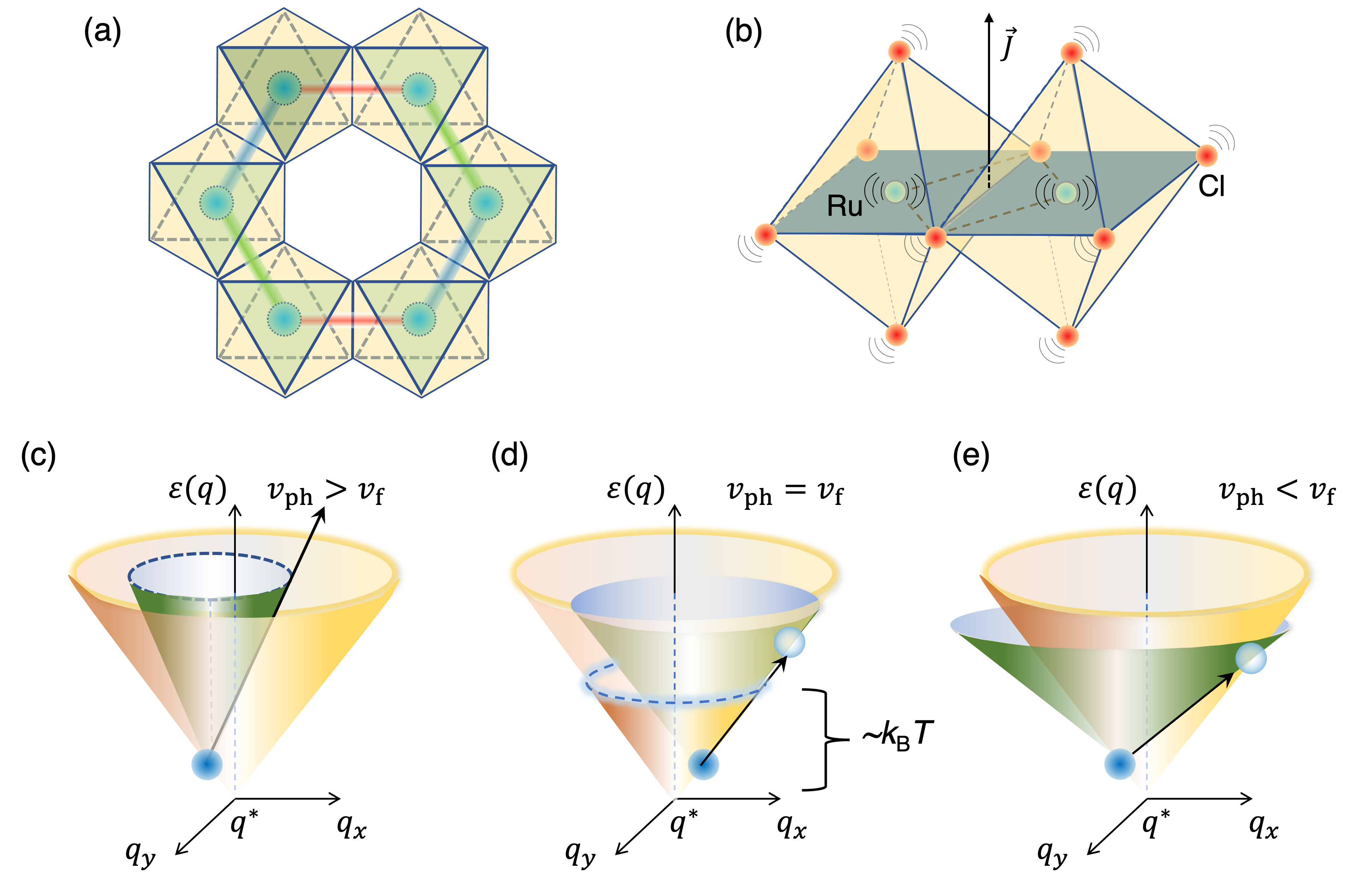}
\caption{Schematics of Majorana-phonon scattering. (a)-(b) Lattice structure of \RuCl{}. The red, blue and green lines in (a) represent the $X$, $Y$ and $Z$ bonds in the Kitaev model (see Eq.~\ref{KitaevModel}). (c)-(e) Schematics of the scattering between phonons (green cone) and MFs (yellow cone) with $v_{\mathrm{ph}}>v_{\mathrm{f}}$, $v_{\mathrm{ph}}=v_{\mathrm{f}}$ and $v_{\mathrm{ph}}<v_{\mathrm{f}}$. }
\label{Fig1}
\end{figure*}

\RuCl{}, consisting of edge-shared Ru-Cl octahedra [Fig.~\ref{Fig1}(a)], is widely believed as a promising Kitaev quantum spin liquid candidate~\cite{Kasahara2018,Yamashita2020,Kitaev2006,Takagi2019,Plumb2014,Sandilands2015,Banerjee2016,Zhou2016,Nasu2016,Banerjee1055,Jansa2018,Widmann2019}. As shown in Fig.~\ref{Fig1}(b), the destructive quantum-interference through the close-to-90$^{\circ}$ Ru–Cl–Ru bonds significantly suppresses the Heisenberg magnetic exchange interaction, yielding a dominant Ising-type interaction ($J_{K}$) perpendicular to the Ru–Cl–Ru plane~\cite{Jackeli2009}. Although non-Kitaev interactions are present in \RuCl{} \cite{Banerjee2016, Banerjee1055, Takagi2019, Laurell2020,Sears2020}, existence of Majorana fermions has been suggested by numerous experimental studies \cite{Sandilands2015, Banerjee2016, Banerjee1055, Do2017, Kasahara2018, Jansa2018, Widmann2019, Yamashita2020, Li2021, Yokoi2021, Czajka2021, Bruin2021}. Most interestingly, a half-integer thermal Hall conductivity in units of $\pi^{2}k_{B}^{2}T/3h$ ($k_{B}$ is the Boltzmann constant) is observed \cite{Kasahara2018, Yamashita2020, Yokoi2021}, supporting non-Abelian anyons in \RuCl{} under in-plane magnetic field. However, the quantized thermal Hall conductivity is found to be sample-dependent \cite{Kasahara2018, Yamashita2020, Yokoi2021,Czajka2021, Bruin2021}, posting questions on the subtle interplay between electronic excitations and lattice degrees of freedom \cite{Lefrancois2021}. In this letter, we combine meV-resolution inelastic x-ray scattering (IXS) and analytical calculation to investigate the Majorana-phonon scattering process. We find that the Majorana-phonon scattering rate diverges when the velocity of acoustic phonons matches the velocity of Majorana fermions. Based on the IXS determined acoustic phonon velocity, we identify the range in the Kitaev interaction, $J_{K}$, for which divergent Majorana-phonon scattering can happen in \RuCl{}, and we find that the two velocities may be very close for reasonable values of $J_{K}$. Our result indicates that structural defects, such as stacking faults, that modify magnetic properties \cite{Banerjee2016, Banerjee1055}, may strongly affect the thermal transport behaviors in \RuCl{}.

Millimeter-sized \RuCl{} crystals were grown by the sublimation of \RuCl{} powder sealed in a quartz tube under vacuum~\cite{May2020}. Magnetic order was confirmed to occur at 7~K by measuring magnetic properties and specific heat~\cite{Banerjee2016}. The IXS experiments were conducted at beam line 30-ID-C (HERIX) at the Advanced Photon Source (APS). The highly monochromatic X-ray beam of incident energy $E_i = 23.7$~keV ($\lambda= 0.5226~\AA$) was focused on the sample with a beam cross section of $\sim 35\times15$~$\mathrm{\mu m}^2$ (horizontal$\times$vertical). The total energy resolution of the monochromatic X-ray beam and analyzer crystals was $\Delta E \sim1.5$~meV (full width at half maximum). The measurements were performed in transmission geometry. Typical counting times were in the range of 30–120 s per point in the energy scans at constant momentum transfer \textbf{Q}. \textbf{H}, \textbf{K}, \textbf{L} are defined in the trigonal structure with $a = b = 5.9639$~$\AA$, $c = 17.17$~$\AA$ at the room temperature.

Before analyzing the IXS result, we analytically derive the velocity dependent Majorana-phonon scattering in a gapless Kitaev spin liquid. Focusing on the lowest temperatures, we consider acoustic phonons of velocity $v_{\mathrm{ph}}$ and linearly dispersing Majorana fermions of velocity $v_{\mathrm{f}}$; the two Dirac points of Majorana fermions at momenta $\pm \mathbf{K}$ (which correspond to the $\mathrm{K}$ and $\mathrm{K}'$ points of the Brillouin zone) are equivalent to a single Dirac point of complex fermions at momentum $\mathbf{K}$. The simplest processes annihilating a phonon of momentum $\mathbf{q}$ then involve the scattering of a fermion with relative momentum $\mathbf{k}$ (with respect to $\mathbf{K}$) into another fermion with relative momentum $\mathbf{k} + \mathbf{q}$. Note that processes creating or annihilating two fermions are forbidden due to the nontrivial momentum $\mathbf{K}$ of the fermion Dirac point ($2 \mathbf{K}$ is not a reciprocal lattice vector). We also remark that, while the phonons are generally three dimensional, only those with a purely in-plane (two-dimensional) momentum $\mathbf{q}$ can participate in the above scattering processes. 

Importantly, due to the conservation of energy and momentum, the Majorana-phonon scattering is strongly velocity dependent. For $v_{\mathrm{ph}} > v_{\mathrm{f}}$ [see Fig.~\ref{Fig1}(c)], a negative-energy fermion can be scattered into a positive-energy fermion, whereas for $v_{\mathrm{ph}} < v_{\mathrm{f}}$ [see Fig.~\ref{Fig1}(e)], a positive-energy fermion can scatter into a positive-energy fermion or a negative-energy fermion can scatter into negative-energy fermion~\cite{Ye2020, Feng2021}. Furthermore, as we show in the Supplemental Material~\cite{SI}, the real and imaginary parts of the phonon self energy (corresponding to the energy renormalization and the decay rate, respectively) both diverge for $v_{\mathrm{ph}} \to v_{\mathrm{f}}$ [see Fig.~\ref{Fig1}(d)]. The divergent components of these two quantities at momentum $\mathbf{q}$ and temperature $T = 1 / \beta$ are found to be
\begin{equation}
    \mathrm{Re} \Sigma_{\mathbf{q}} \propto \bigg\{ \begin{array}{c} +|\mathbf{q}|^2 \left( v_{\mathrm{ph}} - v_{\mathrm{f}} \right)^{-1/2} \mathcal{I}^{(2)} (\beta v_{\mathrm{f}} |\mathbf{q}|) \quad (v_{\mathrm{ph}} > v_{\mathrm{f}}), \\ -|\mathbf{q}|^2 \left( v_{\mathrm{f}} - v_{\mathrm{ph}} \right)^{-1/2} \mathcal{I}^{(1)} (\beta v_{\mathrm{f}} |\mathbf{q}|) \quad (v_{\mathrm{ph}} < v_{\mathrm{f}}), \end{array}
    \label{eq-Re}
\end{equation}
\begin{equation}
    \mathrm{Im} \Sigma_{\mathbf{q}} \propto \bigg\{ \begin{array}{c} |\mathbf{q}|^2 \left( v_{\mathrm{ph}} - v_{\mathrm{f}} \right)^{-1/2} \mathcal{I}^{(1)} (\beta v_{\mathrm{f}} |\mathbf{q}|) \quad (v_{\mathrm{ph}} > v_{\mathrm{f}}), \\ |\mathbf{q}|^2 \left( v_{\mathrm{f}} - v_{\mathrm{ph}} \right)^{-1/2} \mathcal{I}^{(2)} (\beta v_{\mathrm{f}} |\mathbf{q}|) \quad (v_{\mathrm{ph}} < v_{\mathrm{f}}), \end{array}
    \label{eq-Im}
\end{equation}
where $\mathcal{I}^{(1,2)} (x)$ with $x = \beta v_{\mathrm{f}} |\mathbf{q}|$ are positive dimensionless functions with asymptotic forms
\begin{equation}
    \mathcal{I}^{(1)} (x) \sim \bigg\{ \begin{array}{c} x \quad (x \ll 1), \\ 1 \quad (x \gg 1), \end{array} \quad \mathcal{I}^{(2)} (x) \sim \bigg\{ \begin{array}{c} x^{-1} \quad \,\,\,\, (x \ll 1), \\ x^{-3/2} \quad (x \gg 1). \end{array}
    \label{eq-I12}
\end{equation}
Hence, the real and imaginary parts of the phonon self energy have completely different scaling behaviors as a function of $\beta v_{\mathrm{f}} |\mathbf{q}|$. In particular, at the smallest phonon momenta ($\beta v_{\mathrm{f}} |\mathbf{q}| \ll 1$), the divergent contribution is expected to be most observable in the real part for $v_{\mathrm{ph}} > v_{\mathrm{f}}$ and in the imaginary part for $v_{\mathrm{ph}} < v_{\mathrm{f}}$. We also note that the plus (minus) sign in Eq.~\ref{eq-Re} corresponds to hardening (softening) of the phonon energy.

\begin{figure*}
\centering
\includegraphics[width=1\linewidth]{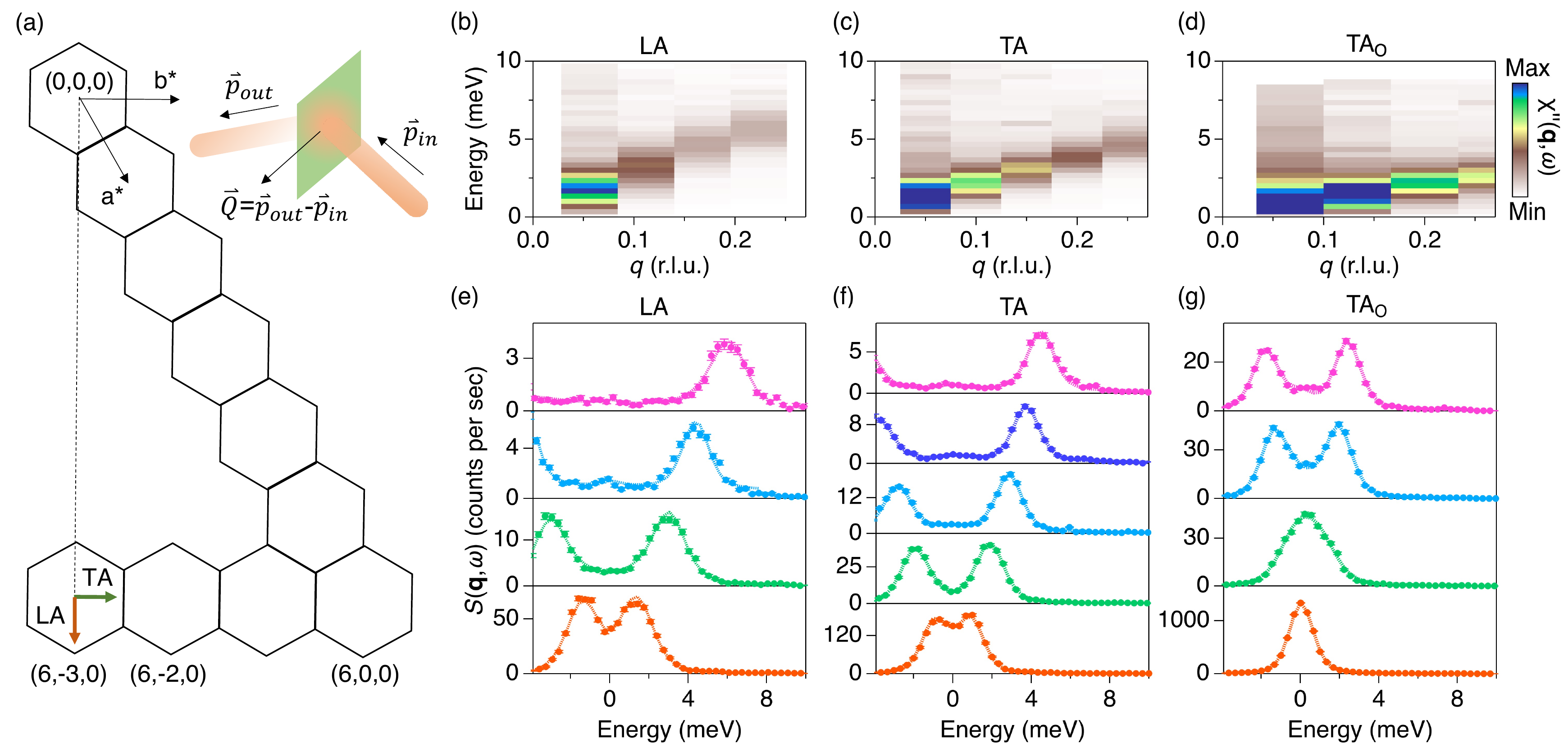}
\caption{Acoustic phonon excitations at small \textbf{q}. (a) Schematics of the IXS experimental setup and trajectory of the measurement in Brillouin zone. (b)-(d) Three branches of acoustic phonon excitations measured by IXS, where LA is measured from the \textbf{Q}=(6, -3, 0)–(6.2, -3.1, 0), TA is measured from \textbf{Q}=(6, -3, 0)–(6, -2.75, 0), TA$_O$ is measured from \textbf{Q}=(6, -3, 0) – (6, -3, 0.266) The plot shows the Bose-factor corrected IXS intensity. (e)-(g)  Raw IXS spectra (solid circles) of the three phonon branches and the fits to the data (dotted lines). The error bars represent one standard deviation assuming Poisson counting statistics. 
}
\label{Fig2}
\end{figure*}

We now determine the in-plane acoustic phonon velocities using IXS. IXS directly probes the phonon dynamical structure factor, $S(\mathbf{Q=q+G},\omega)=\chi\prime\prime(\mathbf{Q},\omega)/(1-e^{-\omega/k_BT})$, where $\mathbf{Q}$ and $\mathbf{q}$ are total and reduced momentum transfer. $\mathbf{G}$ is the reciprocal lattice vector and $\omega$ is the phonon energy. $\chi\prime\prime(\mathbf{Q},\omega)$ is the the imaginary part of the dynamical phonon susceptibility that can be obtained from $S(\mathbf{Q=q+G},\omega)$. For acoustic modes, all atoms in the unit cell move in the same direction as the polarization vector, $\mathbf{e}(\mathbf{q})$, and give rise to $S(\mathbf{Q},\omega)\propto |\mathbf{Q}\cdot\mathbf{e}(\mathbf{q})|^{2}$. Figure~\ref{Fig2}(a) shows the experimental geometry in the trigonal structure at room temperature. To selectively enhance the in-plane longitudinal (LA) and transverse (TA) acoustic phonons, we perform IXS measurement along (6, -3, 0)$\longrightarrow$(7, -3.5, 0) and (6, -3, 0)$\longrightarrow$(6, -2, 0) in reciprocal lattice unit (r.l.u.), respectively. We also performed measurement along (6, -3, 0)$\longrightarrow$(6, -3, 1) to evaluate the out-of-plane transverse acoustic phonon (TA$_o$).

\begin{figure}
\centering
\includegraphics[width=0.5\linewidth]{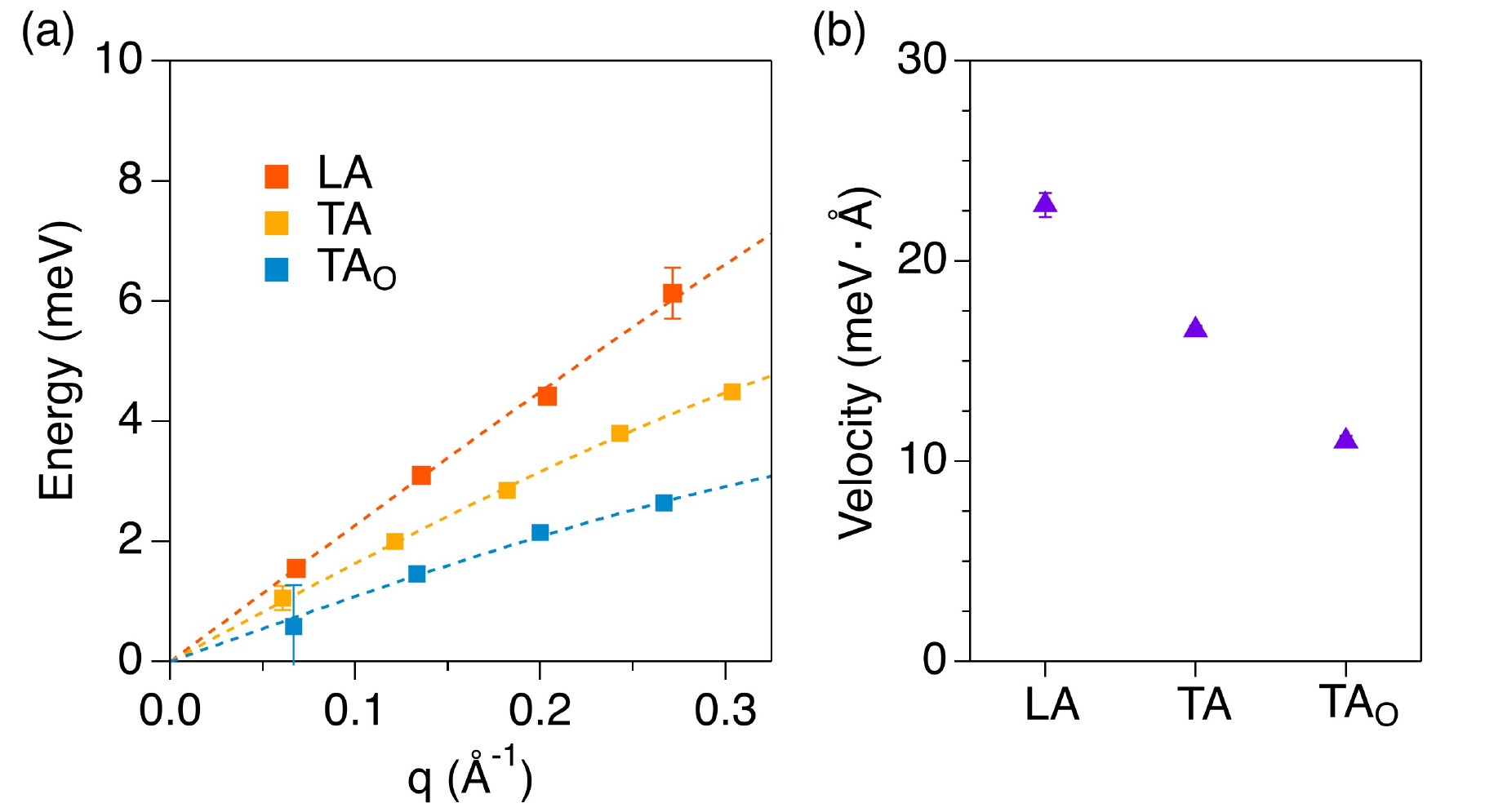}
\caption{Extracted phonon dispersions and velocities. (a) Phonon peak dispersions extracted from the measured $S(\mathbf{Q},\omega)$. The x-axis is in unit of inverse angstrom ($1/\mathrm{\AA}$), thus we can directly compare the dispersion of three acoustic modes. The colored dash line is the sinusoidal fit to the data. (b) The extracted phonon velocities at the $\Gamma$ point based on the dispersions shown in panel (a). Error bars in panel (a) represent the $2\sigma$ returned from spectral peak fittings.  
}
\label{Fig3}
\end{figure}

Figure~\ref{Fig2} shows the image plots of Bose factor corrected dynamical structure factor of in-plane LA [Fig.~\ref{Fig2}(b)], in-plane TA [Fig.~\ref{Fig2}(c)] and out-of-plane TA$_o$ [Fig.~\ref{Fig2}(d)]. Near the Brillouin zone center, the acoustic phonon velocity is defined as $v_{\mathrm{ph}} =\frac{\partial\omega}{\partial q}|_{q\rightarrow0}\sim\frac{\omega}{q}$. In our measurement, the LA mode has the largest slope at the Brillouin zone center and hence the largest $v_{\mathrm{ph}}$. The velocity of TA$_o$ is the smallest at the Brillouin zone center due to the van der Waals bonding between quasi-two dimensional layers. Indeed, at $L=0.5$~r.l.u., the TA$_o$ mode, corresponding to a shear displacement between adjacent \RuCl{}-layers, has an energy of 3~meV at 300~K (see supplemental materials~\cite{SI}), consistent with large stacking fault in \RuCl{}. To precisely extract $\omega(\mathbf{q})$, we fit the raw IXS spectra shown in Figs.~\ref{Fig2}(e)-(g). Phonon modes are represented by standard damped harmonic oscillator functions (DHOFs) \cite{Fak1997}:
\begin{equation}
    \mathrm{DHOF}=\sum_i I_i[\frac{\Gamma_i}{(\omega-\omega_{\mathbf{Q},i})^2+\Gamma_i^2}-\frac{\Gamma_i}{(\omega+\omega_{\mathbf{Q},i})^2+\Gamma_i^2}]
    \label{chidp}
\end{equation}
where $I_i$, $\omega_{\mathbf{Q},i}$, and $2\Gamma_i$ are the intensity, energy, and full width at half maximum (FWHM) of phonon peak $i$. To fit to the IXS spectra, the DHOFs are weighted by the Bose factor and convoluted with the 1.5~meV energy resolution. Details of the spectral peak fitting are given in supplemental materials~\cite{SI}. The extracted phonon peak dispersions are shown in Fig.~\ref{Fig3}(a). The red, orange and blue symbols are corresponding to LA, TA and TA$_o$, respectively. The dashed lines are sinusoidal fittings of the extracted $\omega(\mathbf{q})$.  

As shown in Eqs.~\ref{eq-Re} and \ref{eq-Im}, the Majorana-phonon scattering diverges when $v_{\mathrm{ph}}=v_{\mathrm{f}}$. In the isotropic limit of the pure Kitaev model with $J_K^X=J_K^Y=J_K^Z=J_K$, the Majorana fermion dispersion near the $K$ point is $\omega(\mathbf{k})=\frac{\sqrt{3}}{4}J_K a |\mathbf{k}|$, where the in-plane lattice constant is $a= 5.9639$~\AA. The experimentally determined $v_{\mathrm{ph}}$ can thus be used to identify the range in $J_K$ for which divergent Majorana-phonon scattering can happen. Using the largest acoustic phonon velocity, $v^{\mathrm{LA}}_{\mathrm{ph}}=23$~meV$\cdot$\AA, we obtain $J^{max}_{K}\sim9$~meV. 

Previous experimental and theoretical studies derived a wide range of $|J_K|$ from 1 to 25~meV, although most studies support $5<|J_K|<13$~meV. In the inelastic neutron scattering study, $J_K\sim$ 7~meV was extracted by fitting the spin-wave with a model Hamiltonian \cite{Banerjee2016}, which corresponds to $v_{\mathrm{f}}\sim18$~meV$\cdot$\AA. This value is very close to the phonon velocity of the TA mode with $v^{\mathrm{TA}}_{\mathrm{ph}}=17$~meV$\cdot$\AA, suggesting a close to divergent scattering. Interestingly, a recent IXS study found a 15\% phonon softening in the TA mode at $q\leq$0.1~r.l.u.~\cite{Li2021,SI}. This softening effect is consistent with the Majorana-phonon scattering induced phonon anomaly described by the real part of the phonon self energy in Eq.~\ref{eq-Re}. It is important to emphasize that $J_K\sim$ 7~meV may be incompatible with the high fractionalization temperature, $T^{*}\sim$ 80~K, observed in thermal transport measurements \cite{Kasahara2018}. Numerical study of a pure Kitaev model shows that $T^{*}\sim0.5J_{K}$ \cite{Nasu2016}. For $J_K\sim$ 7~meV, the estimated $T^{*}\sim40$~K is only half of the experimental value. 

According to Eq.~\ref{eq-Im}, the phonon softening will also be accompanied by phonon broadening at low temperature. It is interesting to compare the phonon broadening effect in the electron-phonon coupled charge density wave (CDW) systems, such as the cuprate high-T$_c$ superconductors \cite{LeTacon2011,Miao2018} and transition metal dichalcogenides \cite{Weber2011}. The phonon broadening is usually measured by the damping ratio, $\Gamma/\omega_0$, where $\Gamma$ is the full width at half maximum of the phonon peak, $\omega_0$ is the bare phonon energy. For a 15~\% phonon softening at $\omega_0=2$~meV, as those observed in \RuCl{} at 10~K \cite{Li2021}, the typical damping ratio in the CDW systems are about 0.1 to 0.3 \cite{LeTacon2011, Weber2011, Miao2018}, corresponding to $\Gamma\sim$ 0.2 to 0.6~meV. This broadening effect is beyond the resolving power of the current study (energy resolution $\Delta E\sim$ 1.5~meV). A similar case is the CDW in ZrTe$_3$ \cite{Hoesch2009}, where the phonon broadening effect at the low energy phonon peak $\omega_0\sim2$~meV was not resolved. A sub-meV resolution measurement, such as inelastic neutron scattering, is required to uncover the Majorana-phonon scattering induced phonon broadening.

Finally we discuss the implications of our result to the sample dependent thermal transport measurement in \RuCl{} \cite{Kasahara2018, Yamashita2020, Yokoi2021,Czajka2021, Bruin2021}. Due to the van der Waals structure, the magnetoelastic coupling is expected to be sensitive to material defects, such as stacking faults. Indeed, experimental studies have found that the antiferromagnetic transition temperature can be changed up to 50\% depending on the level of stacking faults. Taking the experimentally estimated $|J_K|$ between 5 and 13~meV, our results indicate that the Majorana-phonon scattering in \RuCl{} may be close to the divergent condition. In this case, the subtle change in structural and/or magnetic interactions can amplify the Majorana-phonon scattering, giving rise to a large impact on their thermal transport properties. 

In summary, we investigated the velocity-dependent Majorana-phonon scattering in the Kitaev quantum spin liquid and found that it shows divergent behavior when the phonon and Majorana velocities match. Our result helps to establish the phonon dynamical structure factor as a promising avenue to uncover fractionalized excitations and sheds light on the novel thermal transport properties of the Kiteav spin liquid candidate \RuCl{}.


Acknowledgement: We thank T. Berlijn, H. Ding, J. K. Keum, G. Kotliar, S. Nagler, N. Perkins, and A. Tennant for stimulating discussions. This research at Oak Ridge National Laboratory (ORNL) was sponsored by the U.S. Department of Energy, Office of Science, Basic Energy Sciences, Materials Sciences and Engineering Division (theory, IXS experiment and material synthesis). This research used resources of the Advanced Photon Source, a U.S. Department of Energy (DOE) Office of Science User Facility, operated for the DOE Office of Science by Argonne National Laboratory under Contract No. DE-AC02-06CH11357. Extraordinary facility operations were supported, in part, by the DOE Office of Science through the National Virtual Biotechnology Laboratory, a consortium of DOE national laboratories focused on the response to COVID-19, with funding provided by the Coronavirus CARES Act. G.B.H. and S.O. were supported by the U.S. Department of Energy, Office of Science, National Quantum Information Science Research Centers, Quantum Science Center.

\appendix

\section{Velocity dependence of the Majorana-phonon scattering}
We consider the low-temperature scattering between acoustic phonons
of velocity $v_{\mathrm{ph}}$ and Majorana fermions of velocity
$v_{\mathrm{f}}$ in a gapless Kitaev spin liquid. The two Dirac
points of Majorana fermions at momenta $\pm \mathbf{K}$ (which
correspond to the $\mathrm{K}$ and $\mathrm{K}'$ points of the
Brillouin zone) are equivalent to a single Dirac point of complex
fermions at momentum $\mathbf{K}$. The simplest processes
annihilating a phonon of momentum $\mathbf{q}$ then involve the
scattering of a fermion with relative momentum $\mathbf{k}$ (with
respect to $\mathbf{K}$) into another fermion with relative momentum
$\mathbf{k} + \mathbf{q}$. Note that the processes creating or
annihilating two fermions are forbidden due to the nontrivial
momentum $\mathbf{K}$ of the fermion Dirac point ($2 \mathbf{K}$ is
not a reciprocal lattice vector). Since the Majorana-phonon coupling
is proportional to the strain $\epsilon_{\mathbf{q}} \sim
|\mathbf{q}| u_{\mathbf{q}}$ with the lattice displacement
$u_{\mathbf{q}} \sim (v_{\mathrm{ph}} |\mathbf{q}|)^{-1/2}
(a_{\mathbf{q}}^{\phantom{\dag}} + a_{-\mathbf{q}}^{\dag})$ in terms
of the bosonic phonon operators $a_{\mathbf{q}}^{\phantom{\dag}}$
and $a_{\mathbf{q}}^{\dag}$, the Majorana-phonon matrix element is
proportional to $(|\mathbf{q}| / v_{\mathrm{ph}})^{1/2}$. Hence, the
imaginary part of the phonon self energy (i.e., the phonon decay
rate) at momentum $\mathbf{q}$, energy $\omega$, and temperature $T$
is given by
\begin{eqnarray}
\mathrm{Im} \Sigma (\mathbf{q}, \omega) &\propto&
\frac{|\mathbf{q}|} {v_{\mathrm{ph}}} \int d^2 \mathbf{k} \, \bigg\{
\left[ \tanh \left( \frac{v_{\mathrm{f}} |\mathbf{k}|} {2T} \right)
+ \tanh \left( \frac{v_{\mathrm{f}} |\mathbf{k} + \mathbf{q}|} {2T}
\right) \right] \delta \big[ \omega - v_{\mathrm{f}} \left(
|\mathbf{k}| + |\mathbf{k} + \mathbf{q}| \right)
\big] \nonumber \\
&& + 2 \left[ \tanh \left( \frac{v_{\mathrm{f}} |\mathbf{k} +
\mathbf{q}|} {2T} \right) - \tanh \left( \frac{v_{\mathrm{f}}
|\mathbf{k}|} {2T} \right) \right] \delta \big[ \omega -
v_{\mathrm{f}} \left( |\mathbf{k} + \mathbf{q}| - |\mathbf{k}|
\right) \big] \bigg\}, \label{eq-Sigma}
\end{eqnarray}
where the first term describes the scattering of negative-energy
fermions into positive-energy fermions, while the second term
accounts for the scattering of positive-energy fermions into
positive-energy fermions and the scattering of negative-energy
fermions into negative-energy fermions. We remark that, while the
phonons may generally be three dimensional, only those with a purely
in-plane (two-dimensional) momentum $\mathbf{q}$ can participate in
the above processes.

Focusing on the first term of Eq.~(\ref{eq-Sigma}), the delta
function $\delta [\omega - v_{\mathrm{f}} (|\mathbf{k}| +
|\mathbf{k} + \mathbf{q}|)]$ necessarily vanishes for $\omega <
v_{\mathrm{f}} |\mathbf{q}|$ due to $|\mathbf{k}| + |\mathbf{k} +
\mathbf{q}| \geq |\mathbf{q}|$. In contrast, for $\omega \geq
v_{\mathrm{f}} |\mathbf{q}|$, the delta function is nonzero along an
ellipse defined by $|\mathbf{k}| + |\mathbf{k} + \mathbf{q}| =
\alpha |\mathbf{q}|$, where $\alpha = \omega / (v_{\mathrm{f}}
|\mathbf{q}|) \geq 1$ is the inverse eccentricity. Taking
$\mathbf{q} = (q,0)$ without loss of generality, and introducing the
elliptical coordinates $\mathbf{k} = q (\cosh \mu \cos \nu - 1,
\sinh \mu \sin \nu) / 2$, this ellipse corresponds to a contour
satisfying $\cosh \mu = \alpha$. Given that $|\mathbf{k}| = q (\cosh
\mu - \cos \nu) / 2$ and $|\mathbf{k} + \mathbf{q}| = q (\cosh \mu +
\cos \nu) / 2$, while $d^2 \mathbf{k} = q^2 d\mu \, d\nu \, (\cosh^2
\mu - \cos^2 \nu) / 4$ and
\begin{equation}
\delta \big[ \omega - v_{\mathrm{f}} \left( |\mathbf{k}| +
|\mathbf{k} + \mathbf{q}| \right) \big] = \delta \left[ q
v_{\mathrm{f}} \left( \alpha - \cosh \mu \right) \right] =
\frac{\delta \left[ \alpha - \cosh \mu \right]} {q v_{\mathrm{f}}} =
\frac{\delta \left[ \mu - \mathrm{arccosh} \, \alpha \right]} {q
v_{\mathrm{f}} |\sinh \mu|} = \frac{\delta \left[ \mu -
\mathrm{arccosh} \, \alpha \right]} {q v_{\mathrm{f}} \sqrt{\alpha^2
- 1}}, \label{eq-delta-1}
\end{equation}
the phonon decay rate from the first term of Eq.~(\ref{eq-Sigma})
then becomes
\begin{eqnarray}
\mathrm{Im} \Sigma^{(1)} (\mathbf{q}, \omega) &\propto& \frac{q^2}
{4 v_{\mathrm{f}} v_{\mathrm{ph}} \sqrt{\alpha^2 - 1}}
\int_{0}^{2\pi} d\nu \left( \alpha^2 - \cos^2 \nu \right) \sum_{\xi
= \pm} \tanh \left[ \frac{q v_{\mathrm{f}} (\alpha + \xi \cos \nu)}
{4T} \right] \nonumber \\
&=& \frac{\sqrt{2} q^3} {v_{\mathrm{f}}} \left( \omega^2 - q^2
v_{\mathrm{f}}^2 \right)^{-1/2} \mathcal{I}^{(1)} \left[
\frac{\omega} {q v_{\mathrm{f}}}, \frac{q v_{\mathrm{f}}} {T}
\right] \quad (\omega > q v_{\mathrm{f}}), \label{eq-Sigma-1}
\end{eqnarray}
where the dimensionless function $\mathcal{I}^{(1)} [\alpha,
\gamma]$ with $\alpha = \omega / (q v_{\mathrm{f}})$ and $\gamma = q
v_{\mathrm{f}} / T$ is given by the integral
\begin{equation}
\mathcal{I}^{(1)} [\alpha, \gamma] = \frac{1} {4 \sqrt{2} \alpha}
\int_{0}^{2\pi} d\nu \left( \alpha^2 - \cos^2 \nu \right) \sum_{\xi
= \pm} \tanh \left[ \frac{\gamma (\alpha + \xi \cos \nu)} {4}
\right]. \label{eq-I-1}
\end{equation}
Since the function $\mathcal{I}^{(1)} [\alpha, \gamma]$ is finite
and well behaved for all $\alpha \geq 1$ and $\gamma > 0$, the
phonon decay rate $\mathrm{Im} \Sigma^{(1)} (\mathbf{q}, \omega)$
from the first term of Eq.~(\ref{eq-Sigma}) is divergent in the
limit of $\alpha = \omega / (q v_{\mathrm{f}}) \to 1$ due to the
factor $(\omega^2 - q^2 v_{\mathrm{f}}^2)^{-1/2}$ in
Eq.~(\ref{eq-Sigma-1}).

Focusing on the second term of Eq.~(\ref{eq-Sigma}), the delta
function $\delta [\omega - v_{\mathrm{f}} (|\mathbf{k} + \mathbf{q}|
- |\mathbf{k}|)]$ necessarily vanishes for $\omega > v_{\mathrm{f}}
|\mathbf{q}|$ due to $|\mathbf{k} + \mathbf{q}| - |\mathbf{k}| \leq
|\mathbf{q}|$. In contrast, for $\omega \leq v_{\mathrm{f}}
|\mathbf{q}|$, the delta function is nonzero along a hyperbola
defined by $|\mathbf{k} + \mathbf{q}| - |\mathbf{k}| = \alpha
|\mathbf{q}|$, where $\alpha = \omega / (v_{\mathrm{f}}
|\mathbf{q}|) \leq 1$ is the inverse eccentricity. Taking
$\mathbf{q} = (q,0)$ without loss of generality, and introducing the
elliptical coordinates $\mathbf{k} = q (\cosh \mu \cos \nu - 1,
\sinh \mu \sin \nu) / 2$, this hyperbola corresponds to a contour
satisfying $\cos \nu = \alpha$. Given that $|\mathbf{k}| = q (\cosh
\mu - \cos \nu) / 2$ and $|\mathbf{k} + \mathbf{q}| = q (\cosh \mu +
\cos \nu) / 2$, while $d^2 \mathbf{k} = q^2 d\mu \, d\nu \, (\cosh^2
\mu - \cos^2 \nu) / 4$ and
\begin{equation}
\delta \big[ \omega - v_{\mathrm{f}} \left( |\mathbf{k} +
\mathbf{q}| - |\mathbf{k}| \right) \big] = \delta \left[ q
v_{\mathrm{f}} \left( \alpha - \cos \nu \right) \right] =
\frac{\delta \left[ \alpha - \cos \nu \right]} {q v_{\mathrm{f}}} =
\frac{\delta \left[ \nu - \mathrm{arccos} \, \alpha \right]} {q
v_{\mathrm{f}} |\sin \nu|} = \frac{\delta \left[ \nu -
\mathrm{arccos} \, \alpha \right]} {q v_{\mathrm{f}} \sqrt{1 -
\alpha^2}}, \label{eq-delta-2}
\end{equation}
the phonon decay rate from the second term of Eq.~(\ref{eq-Sigma})
then becomes
\begin{eqnarray}
\mathrm{Im} \Sigma^{(2)} (\mathbf{q}, \omega) &\propto& \frac{q^2}
{2 v_{\mathrm{f}} v_{\mathrm{ph}} \sqrt{1 - \alpha^2}}
\int_{-\infty}^{\infty} d\mu \left( \cosh^2 \mu - \alpha^2 \right)
\sum_{\xi = \pm} \xi \tanh \left[ \frac{q v_{\mathrm{f}} (\cosh
\mu + \xi \alpha)} {4T} \right] \nonumber \\
&=& \frac{\sqrt{2} q^3} {v_{\mathrm{f}}} \left( q^2 v_{\mathrm{f}}^2
- \omega^2 \right)^{-1/2} \mathcal{I}^{(2)} \left[ \frac{\omega} {q
v_{\mathrm{f}}}, \frac{q v_{\mathrm{f}}} {T} \right] \quad (\omega <
q v_{\mathrm{f}}), \label{eq-Sigma-2}
\end{eqnarray}
where the dimensionless function $\mathcal{I}^{(2)} [\alpha,
\gamma]$ with $\alpha = \omega / (q v_{\mathrm{f}})$ and $\gamma = q
v_{\mathrm{f}} / T$ is given by the integral
\begin{equation}
\mathcal{I}^{(2)} [\alpha, \gamma] = \frac{1} {2 \sqrt{2} \alpha}
\int_{-\infty}^{\infty} d\mu \left( \cosh^2 \mu - \alpha^2 \right)
\sum_{\xi = \pm} \xi \tanh \left[ \frac{\gamma (\cosh \mu + \xi
\alpha)} {4} \right]. \label{eq-I-2}
\end{equation}
Since the function $\mathcal{I}^{(2)} [\alpha, \gamma]$ is finite
and well behaved for all $0 < \alpha \leq 1$ and $\gamma > 0$, the
phonon decay rate $\mathrm{Im} \Sigma^{(2)} (\mathbf{q}, \omega)$
from the second term of Eq.~(\ref{eq-Sigma}) is divergent in the
limit of $\alpha = \omega / (q v_{\mathrm{f}}) \to 1$ due to the
factor $(q^2 v_{\mathrm{f}}^2 - \omega^2)^{-1/2}$ in
Eq.~(\ref{eq-Sigma-2}).

\begin{figure*}
\centering
\includegraphics[width=0.8\linewidth]{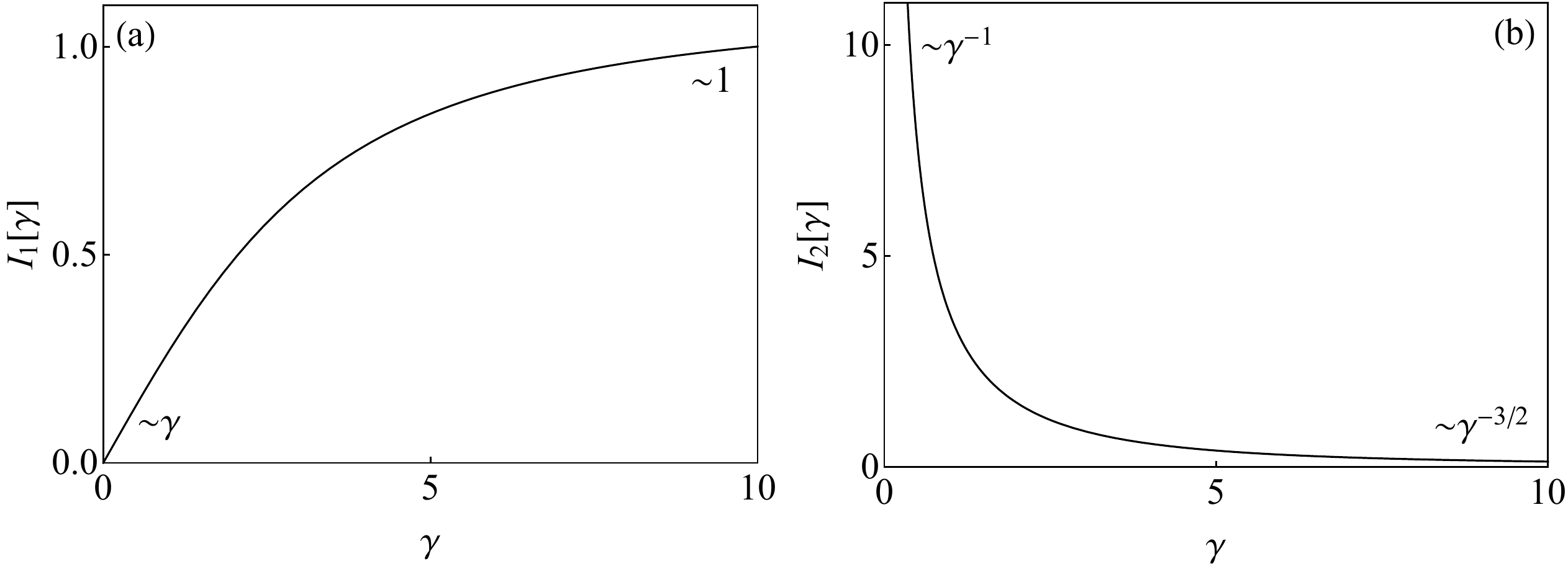}
\renewcommand{\thefigure}{S\arabic{figure}}
\caption{Dimensionless scaling functions $\mathcal{I}^{(1)}
[\gamma]$ and $\mathcal{I}^{(2)} [\gamma]$ (with $\gamma = q
v_{\mathrm{f}} / T \approx q v_{\mathrm{ph}} / T$) that govern the
divergent components of the phonon decay rate and the phonon energy
renormalization in the limit of matching phonon and fermion
velocities ($v_{\mathrm{ph}} \to v_{\mathrm{f}}$).} \label{fig-1}
\end{figure*}

In the limit of $\omega \to q v_{\mathrm{f}}$, the divergent
component of the net phonon decay rate, $\mathrm{Im} \Sigma
(\mathbf{q}, \omega) = \mathrm{Im} \Sigma^{(1)} (\mathbf{q}, \omega)
+ \mathrm{Im} \Sigma^{(2)} (\mathbf{q}, \omega)$, can be extracted
by setting $\alpha = \omega / (q v_{\mathrm{f}}) = 1$ in the first
argument of each dimensionless function $\mathcal{I}^{(1,2)}
[\alpha, \gamma]$. Introducing the relative phonon energy
$\tilde{\omega} = \omega - q v_{\mathrm{f}}$, this divergent
component takes the form
\begin{equation}
\mathrm{Im} \tilde{\Sigma} (\mathbf{q}, \tilde{\omega}) \propto
\sqrt{\frac{q^5} {v_{\mathrm{f}}^3 |\tilde{\omega}|}} \left\{ \theta
\left( \tilde{\omega} \right) \mathcal{I}^{(1)} \bigg[ \frac{q
v_{\mathrm{f}}} {T} \bigg] + \theta \left( -\tilde{\omega} \right)
\mathcal{I}^{(2)} \bigg[ \frac{q v_{\mathrm{f}}} {T} \bigg]
\right\}, \label{eq-Sigma-Im}
\end{equation}
where $\theta (\tilde{\omega})$ is the Heaviside step function,
while the dimensionless functions $\mathcal{I}^{(1)} [\gamma]$ and
$\mathcal{I}^{(2)} [\gamma]$ are given by
\begin{eqnarray}
&& \mathcal{I}^{(1)} [\gamma] \equiv \mathcal{I}^{(1)} [1, \gamma] =
\frac{1} {4 \sqrt{2}} \int_{0}^{2\pi} d\nu \sin^2 \nu \sum_{\xi =
\pm} \tanh \left[ \frac{\gamma (1 + \xi \cos \nu)} {4} \right],
\nonumber \\
&& \mathcal{I}^{(2)} [\gamma] \equiv \mathcal{I}^{(2)} [1, \gamma] =
\frac{1} {2 \sqrt{2}} \int_{-\infty}^{\infty} d\mu \sinh^2 \mu
\sum_{\xi = \pm} \xi \tanh \left[ \frac{\gamma (\cosh \mu + \xi)}
{4} \right]. \label{eq-I-3}
\end{eqnarray}
The precise forms of these functions are plotted in
Fig.~\ref{fig-1}, while their asymptotic behaviors in the limits of
small and large $\gamma$ are
\begin{equation}
\mathcal{I}^{(1)} [\gamma] \sim \Big\{ \begin{array}{c} \gamma \quad
(\gamma \ll 1), \\ 1 \quad (\gamma \gg 1),
\end{array} \qquad \mathcal{I}^{(2)} [\gamma] \sim \Big\{
\begin{array}{c} \gamma^{-1} \quad \,\,\,\, (\gamma \ll 1), \\
\gamma^{-3/2} \quad (\gamma \gg 1).
\end{array} \label{eq-I-4}
\end{equation}
Using the Kramers-Kronig relations, the divergent component of the
phonon energy renormalization, corresponding to the real part of the
phonon self energy, can then be obtained as
\begin{eqnarray}
\mathrm{Re} \tilde{\Sigma} (\mathbf{q}, \tilde{\omega}) &\propto&
\frac{1} {\pi} \, \mathbb{P} \int_{-\infty}^{\infty}
d\tilde{\omega}' \, \frac{\mathrm{Im} \tilde{\Sigma} (\mathbf{q},
\tilde{\omega}')} {\tilde{\omega} - \tilde{\omega}'} = \frac{1}
{\pi} \sqrt{\frac{q^5} {v_{\mathrm{f}}^3}} \Bigg\{ \mathcal{I}^{(1)}
\bigg[ \frac{q v_{\mathrm{f}}} {T} \bigg] \, \mathbb{P}
\int_0^{\infty} \frac{d\tilde{\omega}'} {\sqrt{\tilde{\omega}'}
(\tilde{\omega} - \tilde{\omega}')} + \mathcal{I}^{(2)} \bigg[
\frac{q v_{\mathrm{f}}} {T} \bigg] \, \mathbb{P} \int_0^{\infty}
\frac{d\tilde{\omega}''} {\sqrt{\tilde{\omega}''} (\tilde{\omega} +
\tilde{\omega}'')} \Bigg\} \nonumber \\
&=& \sqrt{\frac{q^5} {v_{\mathrm{f}}^3 |\tilde{\omega}|}} \left\{
\theta \left( \tilde{\omega} \right) \mathcal{I}^{(2)} \bigg[
\frac{q v_{\mathrm{f}}} {T} \bigg] - \theta \left( -\tilde{\omega}
\right) \mathcal{I}^{(1)} \bigg[ \frac{q v_{\mathrm{f}}} {T} \bigg]
\right\}. \label{eq-Sigma-Re}
\end{eqnarray}
Finally, by setting $\omega = q v_{\mathrm{ph}}$ and, hence,
$\tilde{\omega} = q (v_{\mathrm{ph}} - v_{\mathrm{f}})$, the
physical decay rate and energy renormalization are found to be
\begin{eqnarray}
\mathrm{Im} \tilde{\Sigma} (\mathbf{q}) &\propto& \frac{q^2}
{\sqrt{v_{\mathrm{f}}^3 |v_{\mathrm{ph}} - v_{\mathrm{f}}|}} \left\{
\theta \left( v_{\mathrm{ph}} - v_{\mathrm{f}} \right)
\mathcal{I}^{(1)} \bigg[ \frac{q v_{\mathrm{f}}} {T} \bigg] + \theta
\left( v_{\mathrm{f}} - v_{\mathrm{ph}} \right) \mathcal{I}^{(2)}
\bigg[ \frac{q v_{\mathrm{f}}} {T} \bigg] \right\},
\nonumber \\
\mathrm{Re} \tilde{\Sigma} (\mathbf{q}) &\propto& \frac{q^2}
{\sqrt{v_{\mathrm{f}}^3 |v_{\mathrm{ph}} - v_{\mathrm{f}}|}} \left\{
\theta \left( v_{\mathrm{ph}} - v_{\mathrm{f}} \right)
\mathcal{I}^{(2)} \bigg[ \frac{q v_{\mathrm{f}}} {T} \bigg] - \theta
\left( v_{\mathrm{f}} - v_{\mathrm{ph}} \right) \mathcal{I}^{(1)}
\bigg[ \frac{q v_{\mathrm{f}}} {T} \bigg] \right\}.
\label{eq-Sigma-final}
\end{eqnarray}
Both the phonon decay rate and the phonon energy renormalization
diverge in the limit of matching phonon and fermion velocities
($v_{\mathrm{ph}} \to v_{\mathrm{f}}$). Moreover, the divergent
contributions have specific scaling behaviors as a function of
$\gamma = q v_{\mathrm{f}} / T \approx q v_{\mathrm{ph}} / T$ that
can be observed by varying the phonon momentum $\mathbf{q}$ and/or
the temperature $T$. Interestingly, however, there are two
completely different scaling behaviors (see Fig.~\ref{fig-1}), and
the one characterizing the decay rate (energy renormalization) for
$v_{\mathrm{ph}} > v_{\mathrm{f}}$ characterizes the energy
renormalization (decay rate) for $v_{\mathrm{ph}} < v_{\mathrm{f}}$.
Also, the divergent component of the energy renormalization is
positive for $v_{\mathrm{ph}} > v_{\mathrm{f}}$ (corresponding to
phonon hardening) but negative for $v_{\mathrm{ph}} <
v_{\mathrm{f}}$ (corresponding to phonon softening).

\section{Inelastic X-ray scattering and spectral peak fitting}
\begin{figure*}
\centering
\includegraphics[width=0.4\linewidth]{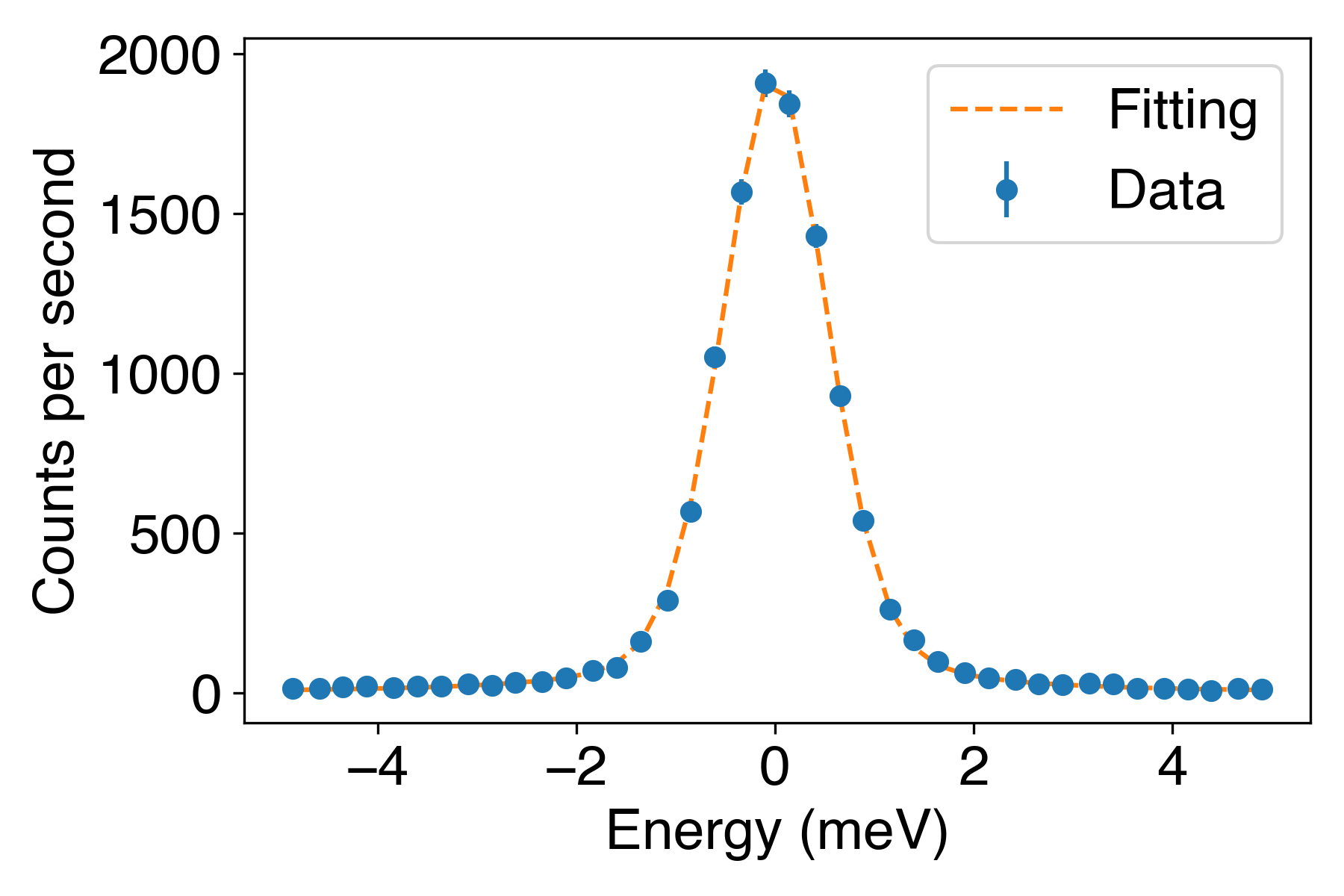}
\renewcommand{\thefigure}{S\arabic{figure}}
\caption{IXS spectrum at Q = (6, -3, 0) r.l.u.. The orange dashed line is a pseudo-voigt function fitting of the experimental data at the room temperature. The extracted energy resolution using the full width half maximum (FWHM) is ~1.3meV. The error bars represent one standard deviation assuming Poisson counting statistics.
}
\label{FigS1}
\end{figure*}

Inelastic X-ray (IXS) scattering is powerful tool for probing phonon excitations with meV energy resolution~\cite{baron2020introduction}. In this work, the experimental setup provides an energy resolution of 1.3 meV and a small X-ray beam size of $\mathrm{35}\times\mathrm{15}$~$\mathrm{\mu m}^2$ . The energy resolution is calibrated by fitting the elastic peak (Fig.~\ref{FigS1}) to a pseudo-voigt function: 
\begin{equation}
    R(\omega)=(1-\alpha)\frac{I}{\sqrt{2\pi}\sigma}e^{-\frac{\omega}{2\sigma^2}}+\alpha\frac{I}{\pi}\frac{\Gamma}{\omega^2+\Gamma^2}  
\end{equation}
where the energy resolution is the full-width-at-half-maximum (FWHM).

IXS directly probes the phonon dynamical structure factor, $S(\mathbf{Q},\omega)$. The IXS cross-section for solid angle $d\Omega$ and bandwidth $d\omega$ can be expressed as \cite{baron2020introduction}:
\begin{equation}
    \frac{d^2\sigma}{d\Omega d\omega}=\frac{k_f}{k_i}r_0^2|\vec{\epsilon_i}\cdot\vec{\epsilon_f}|^2S(\mathbf{Q},\omega)
\end{equation}
where \textbf{k} and $\epsilon$ represent the scattering vector and x-ray polarization and $i$ and $f$ denote initial and final states. $r_0$ is the classical radius of the electron. In a typical measurement, the energy transfer $\omega$ is much smaller than the incident photon energy (23.71 keV in our study). Therefore, the term $\frac{k_f}{k_i}\sim 1$, and $\frac{d^2\sigma}{d\Omega d\omega}\propto S(\mathbf{Q},\omega)$.

$S(\mathbf{Q},\omega)$ is related to the imaginary part of the dynamical susceptibility, $\chi^{\prime\prime}\left(\mathbf{Q},\omega\right)$, through the fluctuation-dissipation theorem~\cite{baron2020introduction}:
\begin{equation}
    S\left(\mathbf{Q},\omega\right)=\frac{1}{\pi}\frac{1}{(1-e^{-\omega/k_BT})}\chi\prime\prime(\mathbf{Q},\omega)
    \label{S(Qw)}
\end{equation}
Where $\chi\prime\prime(\mathbf{Q},\omega)$ can be described by the damped harmonic oscillator form~\cite{baron2020introduction}, which has antisymmetric Lorentzian lineshape:
\begin{equation}
    \chi^{\prime\prime}(\mathbf{Q},\omega)=\sum_i I_i[\frac{\Gamma_i}{(\omega-\omega_{\mathbf{Q},i})^2+\Gamma_i^2}-\frac{\Gamma_i}{(\omega+\omega_{\mathbf{Q},i})^2+\Gamma_i^2}]
    \label{chidp}
\end{equation}
here $i$ indexes the different phonon peaks. 

The phonon peak can be extracted by fitting the IXS spectrum at constant-momentum transfer \textbf{Q}, using Eq.~(\ref{S(Qw)}) and~(\ref{chidp}). Due to the finite experimental resolution, the IXS intensity is a convolution of $S(\mathbf{Q},\omega)$ and the instrumental resolution function, $R(\omega)$: 
\begin{equation}
    I(\mathbf{Q},\omega)=S(\mathbf{Q},\omega)\otimes R(\omega)
    \label{spectralFit}
\end{equation}
Here $R(\omega)$ was determined by fitting elastic peak near a Bragg condition (Fig.~\ref{FigS1}). 

The dashed curves in Fig. 2(e)-(g) of the main text represent fits of the experimental data following Eq. ~(\ref{S(Qw)})~to~(\ref{spectralFit}). In all these fittings, the intrinsic phonon linewidth is much smaller than the energy resolution $\Delta E\sim$1.3~meV. Therefore, all peak-widths are essentially resolution-limited. An additional resolution-limited elastic peak is included to account for the elastic diffuse scattering. The extracted peak positions are shown in Fig. 3(a) of the main text.

\section{Temperature dependence of the in-plane phonon modes}
\begin{figure*}
\centering
\includegraphics[width=0.6\linewidth]{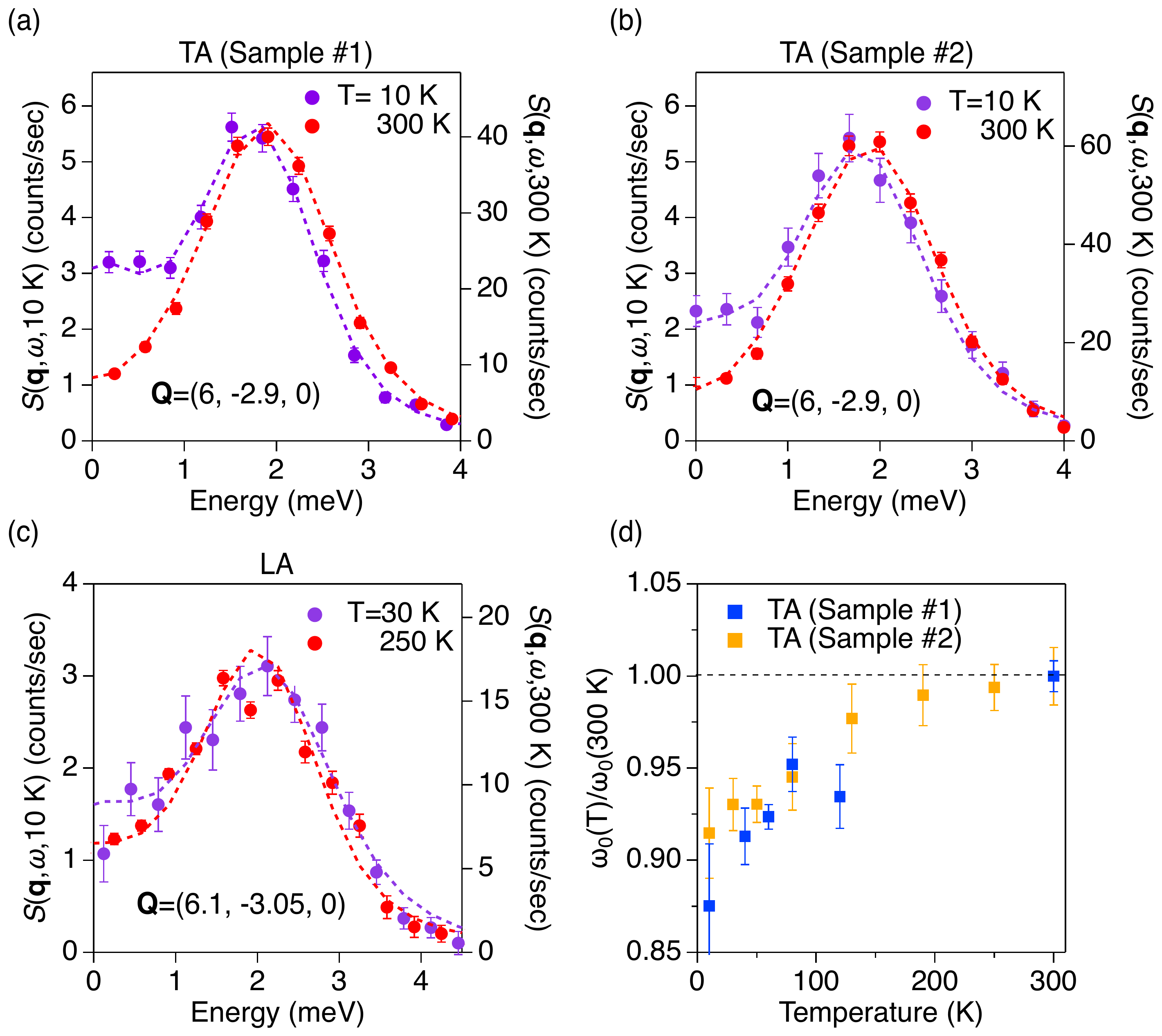}
\renewcommand{\thefigure}{S\arabic{figure}}
\caption{Temperature dependence of the in-plane acoustic phonon modes. (a)-(c) Raw IXS spectra of the transverse acoustic (TA) and longitudinal acoustic (LA) modes at high ($\geq 250$~K) and low temperature ($\leq 30$~K). The dashed lines are the fits to the data. While the TA mode displays a softening effect towards low temperature, the LA mode presents no phonon anomalies. (d) The relative peak shift of TA phonons in sample\#1 and sample\#2.}
\label{FigS2}
\end{figure*}
Figure~\ref{FigS2} present the temperature dependent IXS results on the two in-plane acoustic phonon modes i.e., the transverse acoustic mode (TA) and the longitudinal acoustic mode (LA). Neither the TA mode nor the LA mode display a drastic change over temperature, which demonstrates that the in-plane acoustic phonons maintain the nearly divergent condition down below the Kitaev temperature ($T_K\sim100$~K). In the TA mode, phonon softening effects towards low temperature can be clearly resolved in the raw IXS spectra taken from two different samples [Fig.~\ref{FigS2} (a)(b)], which was reported and discussed in ref.~\cite{Li2021}, whereas the LA mode present no such phonon anomaly. This observation corroborates with the result shown in the main text that the TA mode is closer to the divergent condition compared to the LA mode. 

\section{More on the out-of-plane transverse acoustic phonon branch}
\begin{figure*}
\centering
\includegraphics[width=0.4\linewidth]{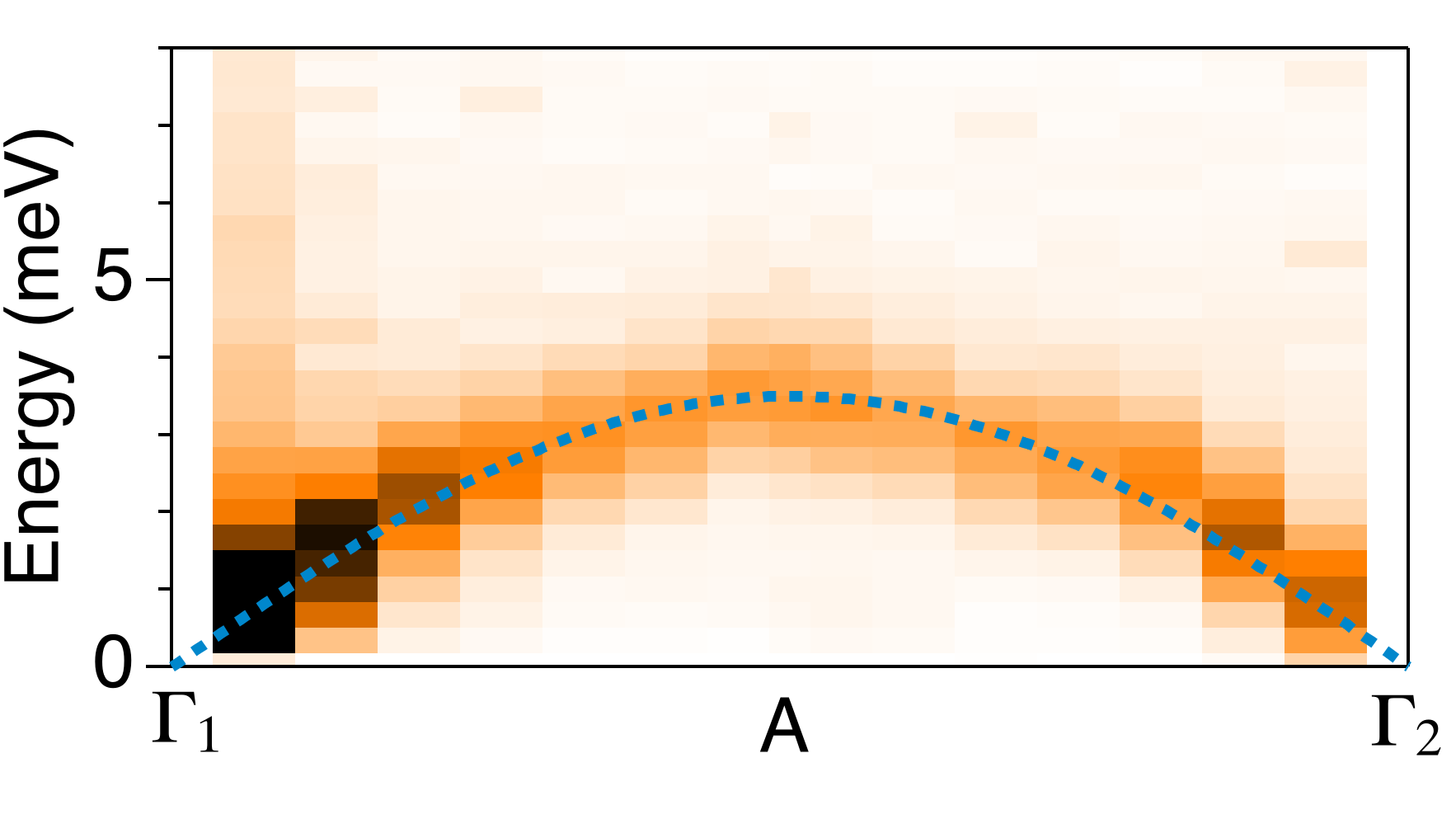}
\renewcommand{\thefigure}{S\arabic{figure}}
\caption{Bose-factor corrected IXS spectrum taken from \textbf{Q}=(6,-3,0) r.l.u. ($\Gamma_1$) to (6,-3,1) r.l.u. ($\Gamma_2$) at $T=300$~K.
}
\label{FigS3}
\end{figure*}
Figure~\ref{FigS3} shows the full dispersion of the out-of plane transverse acoustic phonon (TA$_o$). The TA$_o$ mode, which corresponds to a shear displacement between adjacent \RuCl{} layers, present an extremely low energy band top at only 3~meV. This is consistent with the stacking faults found in this layered material~\cite{jennifer2017,Mi2021}.  

\bibliography{ref}
\end{document}